\documentclass[aps,prd,twocolumn,showpacs]{revtex4}
\newcommand{\M}{\zeta}
\newcommand{\evenM}{{}^{(e)}\!\zeta}
\newcommand{\NevenM}{{{}^{(e)}\!\tilde\zeta}}
\newcommand{\oddM}{{}^{(o)}\!\zeta}
\newcommand{\OevenM}{{}^{(e)}\!\chi}
\newcommand{\OoddM}{{}^{(o)}\!\chi}
\newcommand{\evenpsi}{{}^{(e)}\!\psi}
\newcommand{\oddpsi}{{}^{(o)}\!\psi}
\begin{document}

\title{Improvement on the metric reconstruction scheme in the
  Regge-Wheeler-Zerilli formalism }

\author{Sanjay Jhingan}

\affiliation{Yukawa Institute for Theoretical Physics, Kyoto
University, Kyoto 606-8502, Japan \\ and Dpto. de F\'{\i}sica
Te\'orica, Universidad del Pa\'{\i}s Vasco, Apdo. 644, 48080,
Bilbao, Spain}

\author{Takahiro Tanaka}

\affiliation {Yukawa Institute for Theoretical Physics, Kyoto
  University, Kyoto 606-8502, Japan}

\begin{abstract}
  We study master variables in the Regge-Wheeler-Zerilli
 formalism. We show that a specific choice of new variables is
 suitable for studying perturbation theory from the viewpoint of
 radiation reaction calculations. With explicit definition of the
 improved master variables in terms of components of metric
 perturbations, we present the master equations, with source
 terms, and metric reconstruction formulas. In the scheme using
 these new variables, we do not need any time and radial
 integrations except for solving the master equation. We also show
 that the master variable for even parity modes which satisfies
 the same homogeneous equation as the odd parity case, obtained via
 Chandrasekhar transformation, does not have the good property in
 this sense.
\end{abstract}


\maketitle

\section{Introduction}\label{intro}

Black hole perturbation is a powerful tool for the evaluation of
gravitational waves from a binary system when its mass ratio is
large~\cite{BHp,Takasugi,LoustoPrice}. Although any systematic
method to calculate the radiation reaction to the particle motion
has not been established so far, there are various new
developments in this field
~\cite{MST,Wald,Mino,BO,Lousto,Detweiler02,Barack,NMM,BMNOS,Poisson,MNS,Detweiler,sago,Mino2}.
A formal prescription to extract the self-force was developed
in~\cite{MST}, generalizing the work of DeWitt and
Brehme~\cite{Dewitt} on electromagnetic self-force to include the
gravitational case. These results were further verified by an
independent, and different, axiomatic approach by Quinn and
Wald~\cite{Wald}. The prescription can be summarized as follows.
The retarded field can be obtained in terms of Green's functions
which can be formally decomposed into ``direct'' and ``tail''
parts. Roughly speaking, the ``direct'' part of the field is that
part which has support only on the future light cone, emanating
from the source point. The ``tail" part is composed of
contribution due to curvature scattering which pervades inside the
future light cone of the source point. The analysis presented
in~\cite{MST,Wald} indicates that, the particle motion, after
taking into account the self-force, follows a geodesic on the
geometry perturbed by adding the ``tail" part to the original
background spacetime.

The actual isolation of the ``tail" part is not an easy task.
There are ways to calculate the ``full" Green's function but there
is no direct method to compute the ``tail" part alone. Hence, the
standard prescription that has emerged in the past few years is to
subtract the ``direct" part from the ``full" metric perturbation.
Here lies the well known ``gauge problem". In the standard methods
for constructing full metric perturbation, we first solve the
equation for master variables and then from these master variables
we reconstruct the metric perturbations. The result is naturally
written in a specific gauge such as Regge-Wheeler
(RW)~\cite{Regge,zerilli} or radiation
gauge~\cite{Chrza,Wald2,Ori}. On the other hand, the ``direct"
part is evaluated in the harmonic gauge associated with the
particle trajectory. Therefore, before any meaningful subtraction
we need to relate these expressions which are in different gauges.
This is by no means an easy task since we do not know the
necessary gauge transformation \emph{a priori}. This additional
task to find the appropriate choice of the gauge parameters makes
the problem much harder to solve; this is the aforementioned
``gauge problem". The attempts for subtraction of the ``direct"
part in the RW gauge were reported by Mino  Mino~\cite{Mino} and
Sago et. al.~\cite{sago}.

In this paper we would like to revisit the problem of metric
perturbations reconstruction from the master variables in case of
Schwarzschild background. In this approach, based on
Regge-Wheeler-Zerilli formalism, the problem of metric
reconstruction is relatively well understood~\cite{Regge,zerilli}.
What we would like to discuss here are possibilities of
improvements on this formalism.

It is well known that in the Schwarzschild case the odd and the
even parity perturbations naturally decouple due to spherical
symmetry. Assuming a time dependence of the form $\exp(-i \omega
t)$, the perturbations of a Schwarzschild black hole can be
described by a master equation, for each partial wave mode
decomposed in terms of spherical harmonics, as
\begin{equation}\label{mas-eqn}
\frac{d^{2}\M}{d r^{*2}} + (\omega^{2}-V)\M ={\cal S}(T_{\mu\nu}).
\end{equation}
Here $\M$ is the master variable, ${\cal S}$ is the source term
composed of the matter energy momentum tensor $T_{\mu\nu}$ and
$r^{*}=r+2M\ln(r/2M-1)$ is the usual tortoise radial coordinate.
The metric components in the RW gauge, $h^{RW}$, are obtained by
applying certain differential operators on the master variable and
on the energy-momentum tensor as
\begin{equation}
 h^{RW}=\hat h^{(M)}(\M)
     +\hat h^{(T)}(T_{\mu\nu}).
\end{equation}
Then, the formulas for the metric reconstruction in the scheme
presented in the original papers~\cite{Regge,zerilli} contain
$\omega$ in the denominator of the expressions for $\hat h^{(T)}$.
Although $\omega$ is just a number in the frequency domain, if in
denominator it can be an obstacle in computing the metric in the
vicinity of a particle orbiting a black hole. Suppose that the
particle moves between $r_{min}$ and $r_{max}$. The appearance of
$\omega$ in the denominator means that $\hat h^{(T)}$ is no longer
localized on the radial shell where the particle orbit lies.
Instead, the source is distributed continuously in the region
between $r_{min}$ and $r_{max}$. Therefore, the metric components
are not completely determined by the notion of the master
variables in this region even if the concerned field points are
off the shell. In the computation of the self-force, the
gravitational field exactly on this shell is unnecessary. A
limiting value evaluated along, e.g., the outer radial direction
is sufficient for the purpose of computing the self-force. If we
can modify the formulation so that $\hat h^{(T)}$ is localized on
the shell, then we can apply the formula for the metric
reconstruction outside the source, which is much simpler.  For the
even parity case, an improved master variable has already been
introduced by Moncrief (~\cite{Moncrief}, see also
~\cite{Gerlach}\footnote{We thank Dr. Tomohiro Harada for
informing us about this reference}). We give here the general
metric reconstruction formulas in the presence of sources, which
have not been given explicitly yet, as far as we know. We show
that $\omega$ can be removed from the denominator by using
Moncrief's master variable. Same argument follows for the odd
parity case, i.e., by introducing a new improved master variable,
we can remove appearance of $\omega$ in the denominator. Complete
expressions for the metric reconstruction are also presented for
this case.

Another complication which arises is from the well known fact that
the potentials for odd and even parity cases differ from each
other. The potential for the even parity case is, relatively, much
more complicated. Hence, it would be useful if we could formulate
the even parity perturbations to satisfy the same master equation
with the odd parity case. Chandrasekhar has already given a
unified approach, known as Chandrasekhar transformations, and
shown the relation between RW and Zerilli
equations~\cite{chandra2} (for a comprehensive review
see~\cite{chandra3}). In this paper we also derive the full metric
reconstruction formulas for the even parity perturbation by using
the master variable obtained via the Chandrasekhar transformation.
Under the requirement for this new master variable to satisfy the
RW equation in vacuum, we can still modify its definition by
adding a combination of the metric components which appear on the
left hand side of Einstein equations, since it is zero in vacuum.
Examining all the possibilities of such a modification, we have
concluded that we cannot eliminate $\omega$ from the denominator
in the expression for $\hat h^{(T)}$. Unfortunately, as it turns
out, no dramatic simplification happens by reformulating the
formulas solely in terms of the variable obtained via
Chandrasekhar transformation, although the importance of this
transformation is not reduced at all by this fact.

The paper is organized as follows. In Sec. II we discuss the
equations for both odd and even parity cases. We have provided the
explicit expressions for source terms corresponding to the new
master variables. In Sec. III the even parity master variable
which satisfies odd parity homogeneous master equation is
discussed. We briefly summarize the results obtained in this paper
in Sec. IV, with a speculation towards an alternative method to
compute the regularized self-force subtracting the direct part at
the level of the master variables.

\section{Improved master variables}\label{IMV}
We begin with reexamining the Regge-Wheeler-Zerilli (RW)
formulation. In this formalism a master equation for a master
variable is derived, which are called the RW equation and the RW
variable, respectively. Once we know the solution for the RW
variable, all the components of the metric perturbation can be
derived from it. This scheme is well known. What we have shown
here is that it can be improved, in the sense discussed earlier,
by introducing alternative master variables.

We consider the Schwarzschild metric,
\begin{eqnarray}
 ds^2=&-&\left(1-{2M\over r}\right)dt^2+
 \left(1-{2M\over r}\right)^{-1}dr^2 \nonumber \\
 &+& r^2 \left(d\theta^2+\sin^2\theta \,d\varphi^2 \right)\,,
\end{eqnarray}
as the background. The 10 metric components can be decomposed into
``odd'' and ``even'' parity modes.  We use the notation in which,
after harmonic decomposition for the angular dependence, $H_{0},
H_{1}, H_{2}, h^{(e)}_{0}, h^{(e)}_{1}, K$ and $G$ are the
components of metric perturbations for the even parity modes, and
$h_{0}, h_{1}$ and $h_{2}$ are for the odd parity modes.  Here we
assume that the time dependence is given by $\exp(-i\omega t)$.
Similarly, the components of the energy momentum tensor can be
decomposed into odd and even parity modes.  $A^{(0)}, A^{(1)}, A,
B^{(0)}, B, G^{(s)}$ and $F$ are the expansion coefficients for
the even parity modes, and $Q^{(0)}, Q$ and $D$ are for the odd
parity modes (we follow throughout notation of Zerilli for the
metric perturbation and the energy-momentum tensor with slight
modifications; see~\cite{sago} for the basic equations such as the
law of gauge transformation and the definitions of the harmonic
expansion coefficients of the energy momentum tensor).

\subsection{Odd parity}\label{OP}
First, we consider the odd parity case. The RW gauge choice
corresponds to setting $h_2^{RW}=0$. Here, the variables with a
superscript $RW$ means the quantities are in the RW gauge.  The
nontrivial set of Einstein equations for the odd parity mode is
\begin{eqnarray}\label{odd-ein1}
h^{RW}_{0,rr}+\frac{i\omega}{r^2}(r^2h_{1}^{RW})_{,r}
+\left[\frac{4M}{r}-2(1+\lambda) \right]\frac{h_{0}^{RW}}{r(r-2M)}
\nonumber \\ = \frac{8\pi}{\sqrt{1+\lambda}} \frac{r^2}{r-2M}
  Q^{(0)}\, , \qquad \qquad \qquad \qquad \qquad \\ \label{odd-ein2}
-\omega^2 h_{1}^{RW}+i\omega h_{0,r}^{RW} + 2\lambda
(r-2M)\frac{h_{1}^{RW}}{r^{3}} -i\omega\frac{2}{r}h_{0}^{RW}
\nonumber \\ = -\frac{ 8\pi i}{\sqrt{1+\lambda}} (r-2M) Q\, ,
\qquad \qquad \qquad \qquad \qquad \\ \label{odd-ein3}
\left(1-\frac{2M}{r}\right)
h_{1,r}^{RW}+i\omega\left(1-\frac{2M}{r}
  \right)^{-1}h_{0}^{RW}+\frac{2M}{r^{2}}h_{1}^{RW} \nonumber \\
  =-\frac{8\pi i}{\sqrt{2\lambda(1+\lambda)}} r^2 D\, .
\qquad \qquad \qquad \qquad \qquad  \qquad
\end{eqnarray}
>From the above equations, and using the conventional gauge
invariant master variable $\OoddM$
\begin{equation}\label{Qrw}
\OoddM ={r-2M\over r^2} h_{1}^{RW}\, ,
\end{equation}
we can derive a second order differential equation as
\begin{eqnarray}\label{oddQ}
\left[\partial_{r^*}^2 +\omega^2 -V^{RW}(r)\right] \OoddM
  &=& {\cal S}^{\OoddM} \, .
\end{eqnarray}
This is the well known Regge-Wheeler equation\cite{Regge}.
Here
\begin{equation}
V^{RW}=\left(1-{2M\over r}\right)\left({2(\lambda+1)\over r^2}
-{6M\over r^3}\right)\, ,
\end{equation}
is the Regge-Wheeler potential and the source term is given by
\begin{eqnarray}
{\cal S}^{\OoddM}=
 {8\pi i \over
  \sqrt{\lambda +1}}\left(1-\frac{2M}{r}\right)
 \left[\left(1-\frac{2M}{r}\right) \right.Q \nonumber
 \\ \left. + \frac{r}{\sqrt{2\lambda}}
   \partial_r\left({r-2M\over r} D\right) \right]\, .
   \qquad \qquad
\end{eqnarray}
Here $\lambda$ is defined in terms of $\ell$, the total angular
momentum of the spherical harmonics, as
\begin{equation}
\lambda \equiv \frac{(\ell-1)(\ell+2)}{2} \, .
\end{equation}
Once the master variable and the energy-momentum tensor are given,
we can reconstruct the metric perturbations. To distinguish these
reconstructed variables from the original ones, we associate them
with an over-hat in the same way as $\hat h^{(M)}$ and $\hat
h^{(T)}$. Combining the Einstein equations using the definition of
the master variable, the necessary formulas for the reconstruction
can be derived as
\begin{eqnarray}
{\hat h_1^{RW}}={r^2\over r-2M} \OoddM\,, \qquad \qquad \qquad
 \qquad \cr
{\hat h_0^{RW}} = -{1\over i\omega}\left(1-{2M\over r}\right)
 \left[(r\OoddM)_{,r} \right. \qquad \cr \left. + {8\pi i r^2\over
 \sqrt{2\lambda(\lambda+1)}}D \right]\,. \qquad \qquad \quad
\end{eqnarray}
For $h_1^{RW}$ reconstruction is straight forward
since there is only $\hat h^{(M)}$. In reconstructed
$h_0^{RW}$ first term in the square brackets
corresponds to $\hat h^{(M)}$ and second term is
$\hat h^{(T)}$.  The trouble with the expression for
$\hat h^{(T)}$ is the presence of $\omega$ in
denominator, as anticipated earlier. Even if we
rewrite this expression using conservation law
\begin{equation}
 \sqrt{2\lambda} D = \frac{\omega r^2}{r-2M} Q^{(0)}+
   \left(3-\frac{4M}{r}\right) Q+(r-2M)Q_{,r}\, ,\quad
\label{conserveodd}
\end{equation}
this $\omega$ cannot be removed. This fact implies that we need
time integration of the source term in the reconstruction of
metric perturbation.  Hence the reconstructed metric is not solely
determined by the master variable even when the energy-momentum
tensor vanishes on the spherical shell containing a given field
point.

We, therefore, introduce a new gauge invariant variable $\oddM$
defined by
\begin{equation}\label{new-odd-varRW}
\oddM =-\frac{r}{2 \lambda}\left[-i \omega h^{RW}_{1}-h^{RW}_{0,r}
        +\frac{2}{r}h^{RW}_{0} \right].
\end{equation}
Using the definition (\ref{Qrw}) with one of the odd-parity field
equations, we can verify that
\begin{equation}
 -i \omega \oddM = \OoddM + {8\pi i r(r-2M)Q\over
  {2 \lambda}\sqrt{{1+\lambda}}}\,.
\label{Rtozeta}
\end{equation}
Hence, $\oddM$ is equivalent to {\it time integral} of the
original variable $\OoddM$ outside the source distribution. The
expression applicable to an arbitrary gauge has the same
functional form as in the RW gauge:
\begin{equation}\label{new-odd-var}
\oddM =
-\frac{r}{2 \lambda}\left[-i \omega h_{1}-h_{0,r}+\frac{2}{r}h_{0} \right].
\end{equation}
Substituting Eq. (\ref{Rtozeta}) into Eq. (\ref{oddQ}), we recover
\begin{equation}
\left[\partial_{r^*}^2 +\omega^2 -V^{RW}(r)\right]
 \oddM = {\cal S}^{\oddM},
\label{masterzeta}
\end{equation}
with new source term
\begin{equation}
{\cal S}^{\oddM} = {8\pi (r-2M) \over
  {2 \lambda}\sqrt{1+\lambda}}
 \left[
   \omega r Q
     - \partial_r(r Q^{(0)})\right]\, .
\end{equation}
Here we have used conservation law (\ref{conserveodd}) to simplify
the expression.  The source term ${\cal S}^{\oddM}$ does not have
time integral although $\oddM$ is a time integral of the original
variable $\OoddM$.  This is expected {\it a priori}. If the source
term for $\oddM$ has integration constant then it is not uniquely
determined, which contradicts with the fact that it is a gauge
invariant variable. Here, in order to illustrate the way how we
found the new variable, we took rather lengthy steps to obtain the
master equation (\ref{masterzeta}) passing through the equation
for the original master variable (\ref{oddQ}).  But, of course,
one can directly verify the final result by combining first two
odd-parity equations (\ref{odd-ein1}) and (\ref{odd-ein2}).

Now we consider the reconstruction of the metric components from
this master variable $\oddM$. There are two non-vanishing
components $h_{0}^{RW}$ and $h_{1}^{RW}$ in the RW gauge. They are
to be solely determined from $\oddM$, if the metric perturbation
satisfies Einstein equations.  From Eq.~(\ref{odd-ein2}) and the
definition of $\oddM$, we immediately have
\begin{equation}\label{barh1RW}
\hat h^{RW}_{1}= -\frac{i\omega r^{2}}{r-2M}\oddM +\frac{4\pi i
r^3}{\lambda \sqrt{1+\lambda}} Q \,.
\end{equation}
Once we know $\hat h^{RW}_1$, we can reconstruct $\hat h_0^{RW}$ by
using Eq.~(\ref{odd-ein3}) as
\begin{equation}\label{barh0RW}
\hat h_{0}^{RW}=(r-2M)\left(\oddM_{,r}+\frac{1}{r}\oddM +
\frac{4\pi r^2}{\lambda \sqrt{1+\lambda}}Q^{(0)}\right)\,.
\end{equation}
This time, the $\hat h^{(T)}$ part does not have $\omega$ in the
denominator.  Therefore, one can simply set the source terms to
zero to obtain the formulas for the reconstruction of the metric
perturbation in vacuum region.  We notice here that $\hat h^{(M)}$
is also free from annoying factor $\omega$.  These two facts are
actually related. By definition, the defining expression for a
gauge invariant master variable does not have $\omega$ in the
denominator.  Otherwise, the gauge invariant variable would be
ambiguous due to integration constant, and information of metric
perturbations in the vicinity of a spherical shell, specified by
$t$ and $r$, will be insufficient to determine the corresponding
gauge invariant variable there. Let us assume that $\hat h^{(M)}$
also does not have $\omega$ in the denominator.  In the vacuum
case, we can consider a cycle of operations starting with $h$,
going through the master variable, and again coming back to $h$ by
using $\hat h^{(M)}$.  Throughout this cycle, there is no $\omega$
in the denominator. Hence, if a homogeneous solution of metric
perturbations including its derivatives near a spherical shell is
given, this cycle should reproduce the original metric
perturbations.  The formulas composing this cycle will not change
even if there exists matter source away from the spherical shell.
If $\hat h^{(T)}$ has $\omega$ in the denominator, this term gives
an additional contribution even if matter source does not exist
there. This is a contradiction.  Hence, if $\hat h^{(M)}$ does not
have $\omega$ in the denominator, neither does $\hat h^{(T)}$.

\subsection{even parity}\label{EP}
Next, we look at the even parity case. The RW gauge choice
corresponds to setting $h_0^{(e) RW}=h_1^{(e) RW}=G^{RW}=0$.  The
set of field equations for the even parity modes, with the source
terms, is
\begin{widetext}
\begin{eqnarray}
\label{evena}
\left(1-\frac{2M}{r}\right)\left[\left(1-\frac{2M}{r}\right)
(K^{RW}_{,rr}-\frac{1}{r}
H^{RW}_{2,r})+\left(3-\frac{5M}{r}\right)
\frac{1}{r}K^{RW}_{,r}-\frac{1}{r^2}(H^{RW}_2 -K^{RW})\right.
\qquad\qquad \nonumber
\\\left.-\frac{\lambda}{r^2}(H^{RW}_2
+K^{RW})\right]=-8 \pi A^{(0)}, \\ \label{evenb} -i\omega
K^{RW}_{,r} -i\omega \frac{1}{r}(K^{RW}-H^{RW}_{2})+i\omega
\frac{M}{r(r-2M)}K^{RW}
-\frac{(1+\lambda)}{r^2}H^{RW}_1 =-\frac{8\pi i}{\sqrt{2}}A^{(1)}, &&\\
\label{evenc} \frac{1}{(r-2M)}\left[-\omega^2
\frac{r^2}{(r-2M)}K^{RW}-\left(1-\frac{M}{r}\right)K^{RW}_{,r}+
 2i\omega H^{RW}_{1} +\frac{(r-2M)}{r}H^{RW}_{0,r}
+\frac{1}{r}(H^{RW}_2-K^{RW})\right. \qquad\nonumber
\\\left.+\frac{(1+\lambda)}{r}(K^{RW}-H^{RW}_0)\right]
=-8\pi A,\\ \label{evend}
\left[\left(1-\frac{2M}{r}\right)H^{RW}_1\right]_{,r}+i\omega
(H^{RW}+K^{RW})=\frac{8\pi i}{\sqrt{1+\lambda}} r B^{(0)} ,&& \\
\label{evene} i\omega
H^{RW}_1+\left(1-\frac{2M}{r}\right)(H^{RW}_0-K^{RW})_{,r}+\frac{2M}{r^2}H^{RW}_0
+ \frac{1}{r}\left(1-\frac{M}{r}\right)(H^{RW}_2-H^{RW}_0) =
\frac{8\pi}{\sqrt{1+\lambda}}(r-2M) B, \\
\label{evenf} \omega^2
\left(1-\frac{2M}{r}\right)^{-1}(K^{RW}+H^{RW}_2)+\left(1-
\frac{2M}{r}\right)[K^{RW}_{,rr}-H^{RW}_{0,rr}]+
\left(1-\frac{M}{r}\right)\frac{2}{r}K^{RW}_{,r} -2 i\omega
H^{RW}_{1,r} \qquad \qquad
\nonumber \\
-i\omega
\frac{2(r-M)}{r(r-2M)}H^{RW}_1-\frac{1}{r}\left(1-\frac{M}{r}\right)H^{RW}_{2,r}-
\frac{1}{r}\left(1+\frac{M}{r}\right)H^{RW}_{0,r}-
\frac{(1+\lambda)}{r^2}(H^{RW}_2-H^{RW}_0) = 8\sqrt{2}\pi G^{(s)}
, \\ \label{eveng} H^{RW}_0-H^{RW}_2 = \frac{16
\pi}{\sqrt{2\lambda(1+\lambda)}} r^2 F .
\end{eqnarray}
\end{widetext}
The original Zerilli's master variable, $R_{l m}$, is defined by
\begin{equation}\label{even-var}
R_{l m}=\frac{1}{\omega}\OevenM = \frac{1}{i\omega}
\left(\frac{r-2M}{\lambda r+3M}\right)\left[\frac{i\omega
r^{2}}{r-2M}K^{RW}+H^{RW}_1 \right] ,
\end{equation}
and has an ambiguity due to an integration constant. We will work,
instead, with its gauge invariant form $\OevenM$.  With the field
equations above it obeys the wave equation\cite{zerilli}
\begin{equation}\label{even-eqn}
[\partial^2_{r^*}+\omega^2-V^{Z}(r)]\OevenM={\cal
S}^{(\OevenM)} \, .
\end{equation}
Here,
\begin{eqnarray}\label{old-pot}
V^{Z}(r)=\left(1-\frac{2 M}{r}\right)
\qquad\qquad\qquad\qquad\qquad\qquad \cr \times
\frac{2\lambda^2(\lambda+1)r^3 +6\lambda^2 Mr^2
  +18\lambda M^2 r +18 M^3}{r^3 (r\lambda+3M)^2},
\end{eqnarray}
is the Zerilli potential and the source term takes the form,
\begin{eqnarray}\label{test}
S^{(\OevenM)}=&&\frac{(r-2M)^2}{(r
\lambda+3M)\sqrt{1+\lambda}}B^{(0)}_{,r} \cr
&&+\frac{(r-2M)(-12M^2 + 9Mr + r^2 \lambda)}{r
\sqrt{1+\lambda}(r\lambda+3M)^2}B^{(0)} \cr
&&-\sqrt{2}\lambda\frac{(r-2M)^2}{(r\lambda +3M)^2}A^{(1)}+
\omega\left[-\frac{r^2}{(r\lambda+3M)}A^{(0)}\right. \cr &&
+\frac{(r-2M)^2}{(r\lambda+3M)} A
+\frac{(r+2M)^2}{(r\lambda+3M)\sqrt{1+\lambda}} B \cr && \left. -
\sqrt{2} \frac{(r-2M)}{\sqrt{\lambda (1+\lambda)}} F\right] \, .
\end{eqnarray}

The formulas for the metric reconstruction are derived by combining
the Einstein equations using the definition of the master variable.
Since this is a known result, we just quote here
the explicit reconstruction formula for $K^{RW}$ as an example:
\begin{eqnarray}
{\hat K}^{RW}= \frac{1}{\omega}\left[-\left(1-\frac{2M}{r}\right)
\OevenM_{,r} \right. \qquad\qquad\qquad\cr \left. +
\frac{r^2\lambda+(r\lambda+3M)(r\lambda+2M)}{r^2 (r\lambda+3M)}
\OevenM \right] \cr -\frac{r(r-2M)}{\omega
(r\lambda+3M)}\left(\frac{1}{\sqrt{2}}A^{(1)} +
\frac{1}{\sqrt{1+\lambda}}B^{(0)}\right) \, .
\end{eqnarray}
As in the odd parity case, the first term is $\hat h^{(M)}$ and the
second term is $\hat h^{(T)}$.  The presence of $\omega$ in denominator in
the expression for $\hat h^{(T)}$ is a signal that this $\OevenM$ is
not the most convenient choice of the master variable.

Analogous to the odd parity case we now define a new {\it time
  integrated} variable using vacuum field equations as
\begin{eqnarray}\label{even-par}
\evenM =\frac{r(r-2M)}{(\lambda+1) (\lambda
r+3M)}\left[H^{RW}_2-rK_{,r}^{RW}\right. \qquad\qquad \cr \left.
+\frac{r\lambda+3M}{r-2M}K^{RW} \right]\, .
\end{eqnarray}
In fact, the same variable has been introduced earlier by
Moncrief~\cite{Moncrief} (See also Gleiser et.
al.~\cite{Gleiser}). It can be easily checked that $\evenM$
satisfies a similar wave equation
\begin{equation} \label{even-new-eqn}
[\partial^2_{r^*}+\omega^2-V^{Z}(r)]\evenM={\cal S}^{\evenM},
\end{equation}
with a modified source term
\begin{eqnarray}
\label{even-source}
{\cal S}^{\evenM} = \frac{r-2M}{(1+\lambda)(r\lambda+3M)}
 \left[r^2 A^{(0)}_{,r}-r\left(\frac{r\lambda+2M}{r-2M}
 \right.\right.\cr \left.-\frac{r\lambda+9M}{r\lambda+3M}
  \right)A^{(0)} - \omega \frac{r^2}{\sqrt 2}A^{(1)}+
  (1+\lambda)(r-2M)A \cr
 \left.+\sqrt{1+\lambda} (r-2M)B -
  \sqrt{\frac{2(1+\lambda)}{\lambda}}(r\lambda+3M) F
\right] \, .
\end{eqnarray}
Here, for simplification, we have used the three constraint equations,
corresponding to $T^{\mu}_{~\nu;\mu}=0$, which are
\begin{eqnarray}\label{evencons}
A^{(1)}_{,r}&=&\frac{1}{r-2M}\left[-\frac{\sqrt{2}\omega
r^2}{r-2M} A^{(0)}-2\left(1-\frac{M}{r}\right)A^{(1)} \right. \cr
&&\left. +\sqrt{2(1+\lambda)}B^{(0)}\right],
\\
A_{,r}&=& \frac{1}{r-2M}\left[\frac{\omega r^2}{{\sqrt
2}(r-2M)}A^{(1)} -\frac{M r}{(r-2M)^2}A^{(0)} \right. \cr &&\left.
+\frac{M-2r}{r} A + \sqrt{1+\lambda} B
+\sqrt{2}G^{(s)} \right] ,\\
B_{,r}&=& \frac{1}{r-2M}\left[\frac{\omega r^2}{r-2M}B^{(0)}-
\left(3-\frac{4M}{r}\right) B+ \sqrt{2 \lambda}F \right. \cr &&
\left.- \sqrt{2(1+\lambda)} G^{(s)} \right] .
\end{eqnarray}
As explained in the odd parity case, the source term for a gauge
invariant variable does not have $\omega$ in denominator. Now we
come to the reconstruction of the metric components using this new
master variable $\evenM$ in the RW gauge.  There are four
nonvanishing components in the even parity case, namely, $K^{RW},
H^{RW}_1, H^{RW}_0$ and $H^{RW}_2$. We can rewrite them in terms
of the gauge invariant variable $\evenM$ as
\begin{eqnarray} \label{new-even-var}
\hat K^{RW}&=&\frac{\lambda (\lambda +1)r^2+3\lambda
Mr+6M^2}{r^2(r\lambda+3M)}\evenM
+\left(1-\frac{2M}{r}\right)\evenM_{,r} \cr &&-\frac{8\pi
r^3}{(\lambda +1)(r\lambda + 3 M)}A^{(0)} , \cr
 \hat H^{RW}_1&=& -i\omega
\frac{\lambda r(r-2M)-M(r\lambda +3M)}{(r-2M)(r\lambda +
3M))}\evenM -i\omega r\evenM_{,r} \cr && + i\omega
\frac{r^5}{(1+\lambda )(r\lambda +3M)(r-2M)}A^{(0)} \cr
&&+i\frac{r^2}{{\sqrt 2}(1+\lambda )}A^{(1)}  , \cr {\hat
H^{RW}}_2&=& \frac{1}{r\lambda+3M}\left[\left(-\omega^2r^2
\frac{(r\lambda+3M)}{r-2M}+ \lambda^2 + \frac{3M^2}{r^2} \right.
\right. \cr && \left. +
\frac{\lambda(r^2\lambda+6M^2)}{r(r\lambda+3M)} \right) \evenM-
\left(\frac{M}{r}(r\lambda+3M)-\lambda(2M \right.\cr && \left. -r)
\right)\evenM_{,r}- \frac{1}{1+\lambda} \left\{
\left(\frac{r\lambda}{r\lambda+3M}-
\frac{M}{r-2M}\right)r^3A^{(0)} \right. \cr &&  + \frac{1}{\sqrt
2} \omega r^4 A^{(1)}-(r-2M)r^2[B+ (1+\lambda)A] \cr && \left.+
{\sqrt \frac{2}{\lambda}}r^2(r\lambda+3M)F
\right\}\left. \right], \hspace{-5mm}\nonumber \\
\hat H^{RW}_0 &=& \bar H^{RW}_2 +\frac{16 \pi}{\sqrt{2\lambda(1+\lambda)}} r^2 F .
\end{eqnarray}
These reconstruction formulas are local and do not require any time
integrations.

\section{even parity master variable via Chandrasekhar transformation}
\label{method} In this section we have examined the even parity
master variable that satisfies the same homogeneous master
equation as the odd parity one. The method to obtain such an even
parity master variable is well know as Chandrasekhar
transformation~\cite{chandra2}. Here, we give a short derivation
of this transformation, and discuss the metric reconstruction
scheme using this new variable $\NevenM$. As mentioned earlier, a
part of motivation is the usefulness of master variables which
satisfies the same master equation for both the parities.  In
particular, the master equation is much simpler for odd parity
case. Another point is the appearance of the factor $1/(\lambda
r+3M)$ in the RW potential in the even parity case, which is
absent in the odd parity case.  This factor mathematically means
the existence of a singularity at $r=-3M/\lambda$ in the master
equation.  However, this singularity will not be a physical one
because of the symmetry between even and odd parity cases.  This
factor $1/(\lambda r+3M)$ is inherited in many places of the whole
reconstruction scheme. Although not a serious obstacle in actual
computation, we can expect that the reconstruction scheme might
simplify a lot by using the new variable $\NevenM$.

Our quick derivation of $\NevenM$ is based on the fact that Weyl
scalar contracted with null tetrad $ \psi \equiv -C_{a b c d} l^a
m^b l^c m^d $ satisfies the same homogeneous equation irrespective
of the parity \cite{Teukolsky}.  Here $l^a$ and $m^a$ are outgoing
and angular null tetrad vectors, respectively. For explicit
calculations, we use $ (l^a) =
(r-2M)^{-1}\left(r,r-2M,0,0\right)$, and $(m^a) =
(\sqrt{2}r\sin\theta)^{-1}\left(0,0,\sin\theta,i\right). $ The
following formulas are obtained just by plugging in the explicit
metric form into the definition of $\psi$,
\begin{eqnarray}\label{psiodd}
\oddpsi&=&\frac{i}{2 r^{2} (r-2M)} \left[(r-2M) h_{1,r}^{RW}+ r
h_{0,r}^{RW}-i \omega r h_{1}^{RW}\right. \cr && \left. +\frac{(
2M+i \omega r^{2})}{r-2M}h_{0}^{RW}\right],
\end{eqnarray}
and
\begin{equation}\label{psieven}
\evenpsi=-\frac{(H_{1}^{RW}+H_{2}^{RW})}{r (r-2M)}\, .
\end{equation}
Here the angular dependence, which is given by the spin weighted
spherical harmonics, is suppressed for brevity. We use the same
notation $^{(i)}\psi$ to represent the coefficients of Fourier
harmonic decomposition, but it will not cause any confusion.

Substituting Eqs.~(\ref{barh1RW}) and (\ref{barh0RW}), we can
rewrite Eq.~(\ref{psiodd}) in vacuum as
\begin{eqnarray}
\oddpsi&=&\frac{2}{r^{3} (r-2M)^{2}} [\{ \omega^{2}r^{4}+i \omega
r^{2} (r-3M)\cr &&+(3M-(\lambda +1)r) (r-2M) \}\oddM (r)\cr &&+
r(i \omega r^{2}+3M-r)(r-2M) \oddM_{,r} (r)].
\end{eqnarray}
Here we have used the field equations for simplification.  From the
equation above and with the aid of Eq.~(\ref{masterzeta}) in vacuum, we
can express the master variable $\oddM$ in terms of $\oddpsi$ and its
derivative as,
\begin{eqnarray}\label{rzeta}
\oddM (r)&=&\M[\oddpsi] = \frac{r^{2}}{2(3i \omega M+ \lambda (
\lambda +1))} [\{\omega^2 r^3 - 5i\omega M r \cr &&-4M+r
+\lambda(2M-r) \}\oddpsi(r) - (r-2M)(i\omega r^{2} \cr && +3M-r)
\oddpsi_{,r} (r)].
\end{eqnarray}
Then, with an arbitrary constant ${\cal C}$,
\begin{equation}\label{res1}
\NevenM= {\cal C} \M[\evenpsi]
\end{equation}
should satisfy the RW equation, i.e., the same equation that
$\oddM$ satisfies except for the source term.  After a
straightforward calculation, we obtain
\begin{equation}\label{res2}
\NevenM (r)=2(r-2M)\left(H_{2}^{RW}-rK^{RW}_{,r}+
\frac{r\lambda}{(r-2M)}K^{RW} \right),
\end{equation}
with
\begin{equation}\label{star}
{\cal C }=4(3i\omega M+\lambda (1+\lambda)),
\end{equation}
which is known as the Starobinsky constant\cite{chandra3}.  It is
also easy to check directly that this new master variable
satisfies the homogeneous RW equation with the same $V$ as the
usual RW potential.

In general case with the source term, we have
\begin{widetext}
\begin{equation}\label{inmaster}
[\partial_{r^{*}}^{2}+\omega^{2}-V^{RW}(r)] \NevenM
   ={\cal S}^\NevenM,
\end{equation}
with
\begin{eqnarray}\label{res-source}
S^\NevenM
= (r-2M)& & \left[ -2r A^{(0)}_{,r}+2\frac{M-r(1+\lambda)}{r-2M} A^{(0)}
+\sqrt{2}\omega r A^{(1)}+ 2\frac{[6M-r(1+\lambda)](r-2M)}{r^2} A
\right. \nonumber \\
& &+2\frac{[6M-r(1+\lambda)](r-2M)}{\sqrt{1+\lambda}r^2}B
-\left. 2\sqrt{2}{\frac{[6M^{2}-\lambda {r}^{2}
(1+\lambda) ]}{\sqrt{(1+\lambda)\lambda r^2}}}F
-6\sqrt{2}{\frac{rM(r-2M)}{\sqrt{(1+\lambda)\lambda r}}}F_{,r} \right].
\end{eqnarray}

The metric reconstruction formulas for the $\NevenM$ are given by
\begin{eqnarray}
\hat K&=&\frac{16}{|{\cal C}|^2}
 \left[-r(1+\lambda )(r\lambda+3M ) A^{(0)}
+\frac{3}{r} \sqrt{1+\lambda} M(r-2M)^{2}\{B+\sqrt{1+\lambda}
A\} -\frac{3}{\sqrt{2}}\omega r M(r-2M ) A^{(1)} \right.
\nonumber \\ &&-
 3\frac{\sqrt{2(1+\lambda )}M(r-2M)(r\lambda+3M )}{r\sqrt\lambda}F +
 \frac{[(1+\lambda)\{3M(r-2M)+r\lambda(r\lambda+3M)\}-r{\cal O}]}{2r^3}
  \NevenM \nonumber \\&&+\left.
 \frac{(r\lambda +3M)(r-2M)(1+\lambda)}{2r^2} \NevenM_{,r}
  \right] , \nonumber \\
\hat H_1&=&\frac{16}{|{\cal C}|^2} \left[-i\omega
\frac{{\cal P}r^2( r\lambda+3 M)}{(r-2 M)}A^{(0)} +i \frac{[
3 M{\omega}^{2}r +{\lambda}^2(\lambda+1)]r^2}{\sqrt{2}} A^{(1)}
+3 i\omega M {\cal P} (r-2 M)\{A + \frac{B}{\sqrt{1+\lambda}} \}\right.
\nonumber \\
&&- \left.3\sqrt{2}i\omega\frac{M{\cal P}(r\lambda+3
M)}{ \sqrt {(1+\lambda) \lambda}}F +i\omega \frac{[r^2{\cal O} +
3M{\cal P}(r-2M)]}{2{r}^{2} (r-2 M) }\NevenM
+ i\omega \frac{( r\lambda+3M ){{\cal P}}}{2r} \NevenM_{,r} \right],
\nonumber\\
\hat H_2&=&\frac{16}{|{\cal C}|^2} \left[
\frac{r^2 {\cal O}}{(r-2 M )} A^{(0)} + \omega\frac{\lambda {r}^{2}
{\cal P}}{\sqrt{2}} A^{(1)} - \lambda (r-2M){\cal P}\{(1+\lambda)
A + \sqrt{1+\lambda} B\} +\sqrt{2( 1+\lambda ) \lambda} {\cal P}( r\lambda+3 M)
 F \right. \nonumber \\ &&+ \left.  \frac{[{\cal O}-
 \lambda(\omega^2r^3+M(1+\lambda))]{\cal P}}{2
 (r-2M) r^2} \NevenM - \frac{{\cal O}}{2 r}
 \NevenM_{,r} \right], \nonumber \\
\hat H_0 &=& \bar H_2 +\frac{16 \pi}{\sqrt{2\lambda(1+\lambda)}} r^2 F ..
\end{eqnarray}
\end{widetext}
Here, ${\cal P}=3M-r(1+\lambda)$ and ${\cal
O}=3M\omega^2r^2+ \lambda(\lambda+1)(3M-r)$.  If we are working in
frequency domain only, the above choice of master variable is not
a bad one because of the common potential in master equation.
Whereas, in the time domain we will need time integrations for the
metric reconstruction due to the factor $|{\cal C}|^{-2}$.

We can modify the master variable by adding combinations of metric
components which appears on the left hand side of the Einstein
equations.  Let's denote these combinations by $G_{(i)}$ so that the
Einstein equations are formally written as $G_{(i)}=T_{(i)}~(i=1,
2,\cdots, 7)$, where $T_{(i)}$ represents each component of the energy
momentum tensor, $\{A^{(1)}, A, B, A^{(0)}, B^{(0)}, G^{(s)}, F\}$.
Since $G_{(i)}$ vanishes identically outside the source, the
homogeneous equation for the modified master variable should be
unaltered by the transformation,
\begin{equation}
  \NevenM \to \NevenM +\sum_{i=1}^{7} c_i\, G_{(i)} \, .
\end{equation}
Now one may think that the factor $|{\cal C}|^{-2}$ from the
expression for $\hat h^{(T)}$ can be eliminated by using this degree
of freedom of modifying the master variable. However, we will prove
below that it is impossible.

As a result of the transformation above, $\hat K^{(T)}$ is modified as
\begin{equation}
  \hat K^{(T)} \to \hat K^{(T)}-\hat K^{(M)}
   \left[\sum_{i=1}^{7} c_i\, T_{(i)}\right].
\end{equation}
Since $\hat K^{(M)}[\NevenM]$ contains $\NevenM_{,r}$, we cannot
eliminate $T_{(i),r}$ from $\hat K^{(M)}[\sum c_i\, T_{(i)}]$ unless
$c_i = 0$ for $i\geq 4$.  For $\{A^{(1)}, A, B\}$, one can use the
conservation law (\ref{evencons}) to eliminate $T_{(i),r}$.  Thus, the
condition that $|{\cal C}|^2 \hat K^{(M)}[\sum c_i\, T_{(i)}]\approx
0$ requires $c_i \approx 0$ for $i\geq 4$, where $\approx$ means the
equality modulo $|{\cal C}|^2$.  Then, we find that $B^{(0)}$ and
$G^{(s)}$ arises in the expression for the modified $\hat K^{(T)}$
only from $\sum_{i\leq3} c_i\, T_{(i),r}$.  Hence, the conditions for
the coefficients of $B^{(0)}$ and $G^{(s)}$ to vanish modulo $|{\cal
  C}|^2$ become $c_1\approx -(\omega r^2/
\sqrt{2(1+\lambda)}(r-2M))c_3$, and $c_2\approx \sqrt{(1+\lambda)}
c_3$, respectively.  Thus a possible modification which might
eliminate the factor $|{\cal C}|^{-2}$ from the expression for $\hat
h^{(T)}$ is restricted to
\begin{eqnarray}
   \hat K^{(T)} &\to& \hat K^{(T)}-\hat K^{(M)}
  [f(r) ( -(\omega r^2/ \sqrt{2(1+\lambda)}(r-2M)) \cr && \times
  A^{(1)}+\sqrt{(1+\lambda)}A+B)],
\end{eqnarray}
with an arbitrary function $f(r)$. Then, a straightforward
calculation shows that the factor $|{\cal C}|^{-2}$ cannot be
eliminated by this transformation.  Thus, the idea of introducing
a new master variable for even parity modes satisfying the RW
equation does not work well for the purpose of metric
reconstruction in time domain.

\section{discussion}\label{discussion}
In this paper we have introduced new master variables for the odd
and the even parity cases. We call them, respectively, the
modified Regge-Wheeler and Zerilli variables. These variables
satisfy the same Regge-Wheeler or Zerilli equation except for the
source terms, which are composed of the matter energy momentum
tensor. We have given the explicit expressions for the source
term. The metric perturbation in the RW gauge is expressed in
terms of the master variables and the matter energy momentum
tensor. The explicit formulas for the metric reconstruction were
also written down. The important aspect of these modified
variables lies in the fact that the frequency $\omega$ does not
appear in the denominator in all the formulas to obtain the metric
perturbation. Hence, there is no time integration except for the
step solving the master equation. The most crucial point will be
that $\hat h^{(T)}$, the contribution to the reconstructed metric
perturbation from the matter energy-momentum tensor, does not have
$\omega$ in the denominator. Therefore, the perturbed metric
around a field point $(t,r)$ is solely written in terms of the
master variables if the energy-momentum tensor vanishes in the
vicinity of the spherical shell containing this field point. This
fact will be useful in the program to calculate the regularized
self-force acting on a particle orbiting in the Schwarzschild
spacetime.

As mentioned earlier, in the Introduction, the full metric
perturbation contains a divergent piece near the particle
location. To obtain a sensible expression for the self-force, we
need to subtract the so-called ``direct" part from the full metric
perturbation before evaluating the expression of the force. But
here is the ``gauge problem". The full metric perturbation is
obtained in Regge-Wheeler gauge but the ``direct" part is
evaluated in the harmonic gauge associated with the particle
trajectory.

Here we would like to propose an insight towards an alternative
method to handle this gauge issue in the case of the Schwarzschild
background. The basic idea is inspired by the notion brought by
Barack and Ori \cite{BO}. They stressed that the trajectory in the
perturbed spacetime is gauge invariant although the expression for
the self-force depends on the choice of gauge. On the other hand,
the metric perturbations reconstructed from this gauge invariant
master variables depend on the choice of gauge, but the concepts
of the perturbed geometry and hence of the geodesic on it are
gauge invariant. Hence, naturally one may expect that the
subtraction at the level of gauge invariant master variables is
possible.

A sketch of the new method is the following. The ``direct'' part
of the metric perturbation $h^{(S)}$ can be calculated in the
harmonic gauge. We can use the recent observation by Detweiler and
Whiting \cite{Detweiler} that $h^{(S)}$ can be modified so that it
satisfies the Einstein equations. Since the method for the
harmonic decomposition of the direct part is established by Mino,
Nakano and Sasaki\cite{MNS}, the projection of this direct part to
the gauge invariant master variable $\zeta^{(S)}$ is possible by
using the formulas (\ref{new-odd-var}) and (\ref{even-var}).  On
the other hand, solving the RW equation, we can directly calculate
the master variable corresponding to the full metric perturbation,
$\zeta^{(full)}$.  Then we subtract the direct part $\zeta^{(S)}$
from $\zeta^{(full)}$ to obtain the master variable that
corresponds to the tail part, which we denote by $\zeta^{(tail)}$.
Since both $\zeta^{(S)}$ and $\zeta^{(full)}$ satisfy the RW
equation with the same source, their difference $\zeta^{(tail)}$
satisfies the homogeneous RW equation. Hence, we can reconstruct
the metric perturbation corresponding to the tail part from this
regularized master variable $\zeta^{(tail)}$ by applying the
formulas $\hat h^{(M)}$.  At this step the choice of gauge is
unimportant as is explained in the paper by Barack and Ori
\cite{BO}.  Since the subtraction of the divergent part is done at
the level of the gauge invariant variables, we would like to call
this scheme the {\em gauge invariant regularization}.

In the new scheme, using the variables introduced in this paper,
the part depending on the master variable in the metric
reconstruction formulas, $\hat h^{(M)}$, does not have $\omega$ in
denominator as well as $\hat h^{(T)}$ .  Hence, when we know the
behavior of the master variable corresponding to a homogeneous
solution of metric perturbations in the vicinity of a spherical
shell, we can reproduce the metric perturbations from the master
variable there.  If $\hat h^{(M)}$ contained $\omega$ in
denominator, the local information of the master variables near
the shell were not sufficient to reproduce the metric
perturbations.  Therefore, the use of the new variables introduced
in this paper is crucial for the gauge invariant regularization.

This scheme still has a subtle point which requires further
investigation. The method for the reconstruction of the metric
perturbation does apply only for a solution of vacuum Einstein
equations. However, in the actual computation, the direct part
$h^{(S)}$ is calculated in a power series expansion with respect to
the separation $\xi$ between the source point and the field point, and
this expansion must be truncated at a certain order of $\xi$. Then, the
truncated direct part does not satisfy the Einstein equations in
general.  Hence, we need a new invention to bypass this difficulty in
order to realize this attractive idea of the gauge invariant
regularization. We would like to return to this challenging issue in
a future publication.

In Sec. III we discussed the possibility of using a master
variable for even parity modes which has the same potential for
the master equation as in the case of odd parity modes. Such a
variable is obtained by using the Chandrasekhar transformation. We
wrote down the explicit definition of this master variable in
terms of the metric components, the master equation with the
source terms and the metric reconstruction formulas. We found that
the metric reconstruction formulas {\it necessarily} contain the
Starobinsky constant including $\omega$ in the denominator.
Therefore, the use of the even parity master variable that has the
same homogeneous master equation as in the odd parity case
unfortunately turned out not to be advantageous. However, the
master variables considered here are limited to those which are
related via Chandrasekhar transformation. We expect an even wider
class of transformations in which we might find a more suitable
variable for the purpose of metric reconstruction.

\acknowledgments We would like to express special thanks Y. Mino who
gave lot of insight throughout this work. It is also our pleasure to
acknowledge H. Nakano, M. Sasaki, N. Sago and participants of the
workshop YITP-W-01-16 held at Yukawa Institute, the Radiation Reaction
Focus Session and the 5$^{th}$ Capra Ranch meeting at Penn State, for
useful discussions related to this work.  SJ acknowledges support
under COE program and Basque Govt. fellowship.  This work is suppored
in part by Grant in aid for Scientific Research of Japanese Ministry
of Education, Culture, Sports, Science and Technology No. 14047212.

\end{document}